\documentclass[useAMS, usenatbib]{mn2e}

\usepackage{epsfig}

\usepackage{amssymb}

\usepackage{multirow}

\usepackage{enumitem}

\usepackage{ulem}

\usepackage{rotating}
\usepackage{natbib}
\usepackage{latexsym}
\bibpunct{(}{)}{;}{a}{}{,}

\usepackage[hyphens]{url}

\begin{document}

\newcommand{\ledd}{%
$L_\mathrm{Edd}$}

\newcommand{\fgl}{3FGL~J0212.1+5320}

\newcommand{\Msun}{M$_\mathrm{\odot}$}

\def\rem#1{{\bf #1}}
\def\hide#1{}

\def \aj {AJ}
\def \mnras {MNRAS}
\def \apj {ApJ}
\def \apjs {ApJS}
\def \apjl {ApJL}
\def \aap {A\&A}
\def \aapr {A\&ARv}
\def \nat {Nature}
\def \araa {ARAA}
\def \pasp {PASP}
\def \aaps {AAPS}
\def \prd {PhRvD}
\def \apss {ApSS}

\newcommand{\specialcell}[2][c]{%
  \begin{tabular}[#1]{@{}c@{}}#2\end{tabular}}

\author[Linares et al.]{
\parbox[t]{\textwidth}{
\raggedright 
Manuel Linares$^{1,2,3}$\thanks{Linares@iac.es},
Paulo Miles-P{\'a}ez$^{1,2}$,
Pablo Rodr{\'i}guez-Gil$^{1,2}$,
Tariq Shahbaz$^{1,2}$,
Jorge Casares$^{1,2,4}$,
Cecilia Fari\~na$^{1,5}$,
Raine Karjalainen$^{5}$
}
\vspace*{4pt}\\
$^1$ Instituto de Astrof{\'i}sica de Canarias, c/ V{\'i}a L{\'a}ctea
s/n, E-38205 La Laguna, Tenerife, Spain\\
$^2$ Universidad de La Laguna, Departamento de Astrof{\'i}sica,
E-38206 La Laguna, Tenerife, Spain\\
$^3$ Institutt for fysikk, NTNU, Trondheim, Norway\\
$^4$ Department of Physics, Astrophysics, University of Oxford, Denys
Wilkinson Building, Keble Road, Oxford OX1 3RH, UK\\
$^5$ Isaac Newton Group of Telescopes, E-38700 S.C. de La Palma, Tenerife, Spain\\
}

 \title[A new redback MSP]{A millisecond pulsar candidate in a 21-hr orbit: \fgl}

\maketitle{}

\begin{abstract}

We present the discovery of a variable optical counterpart to the
unidentified gamma-ray source \fgl\, and argue this is a new compact
binary millisecond pulsar (MSP) candidate.
We show \fgl\ hosts a semi-detached binary with a
0.86955$\pm$0.00015~d orbital period and a F6-type companion star at
an estimated distance of D=1.1$\pm$0.2~kpc, with a radial velocity
curve semi-amplitude K$_2$=214.1$\pm$5.0~km~s$^{-1}$ and a projected
rotational velocity of Vsin(i)=73.2$\pm$1.6~km~s$^{-1}$.
We find a hard X-ray source at the same location with a 0.5--10~keV
luminosity
L$_\mathrm{X}$=2.6$\times$10$^{32}$~(D/1.1~kpc)$^2$~erg~s$^{-1}$,
which strengthens the MSP identification.
Our results imply a mass ratio q=M$_2$/M$_1$=0.26$^{+0.02}_{-0.03}$ if
the companion star fills its Roche lobe, and q$\gtrsim$0.26 in any case.
This classifies \fgl\ as a ``redback'' binary MSP; if its MSP nature
is confirmed, this will be the brightest compact binary MSP in the
optical band (r'$\simeq$14.3~mag) and will have the longest orbital
period among Galactic field systems (nearly 21~hr).
Based on the light curve peak-to-peak amplitude ($\Delta$r=0.19~mag), we
further suggest that the orbital inclination is high
and the putative pulsar mass is close to canonical
(M$_1$$\simeq$1.3--1.6~M$_\odot$).
Finally, we discuss the lack of heating signatures and asymmetric
optical light curves in the context of other redback MSPs.

\end{abstract}


\vspace{-0.5cm}

\section{Introduction}
\label{sec:intro}

Nearly one thousand gamma-ray sources from the {\it Fermi} Large Area
Telescope (LAT) catalog remain unindentified, about a third of the
total sample \citep{Acero15}.
This is often due to the lack of counterparts at longer wavelengths,
and offers an appealing discovery space.
Among the identified Galactic sources, pulsars are the most numerous
class \citep{Nolan12,Abdo13}, and {\it Fermi}-LAT is uncovering a new
population of nearby binary millisecond pulsars \citep[MSPs; see,
  e.g.,][]{Hessels11,Ray12,Roberts13}.

Dynamical studies of a few MSPs in compact binaries (``black-widow''
and ``redback'' pulsars) have revealed evidence for massive neutron
stars, with masses well above the
% once 
canonical value of 1.4~M$_\odot$
\citep{Kerkwijk11,Romani12b,Kaplan13}.
These and related pulsar discoveries have pushed the maximum neutron
star mass to more than two solar masses
\citep{Demorest10,Antoniadis13}, placing tighter constraints on the
equation of state above nuclear saturation density.
Finding more such systems is crucial to establish their properties as
a class, and constitutes a promising first step towards identifying
the most massive neutron stars.

\begin{figure*}
%\centering
  \begin{center}
  \resizebox{1.7\columnwidth}{!}{\rotatebox{-90}{\includegraphics[]{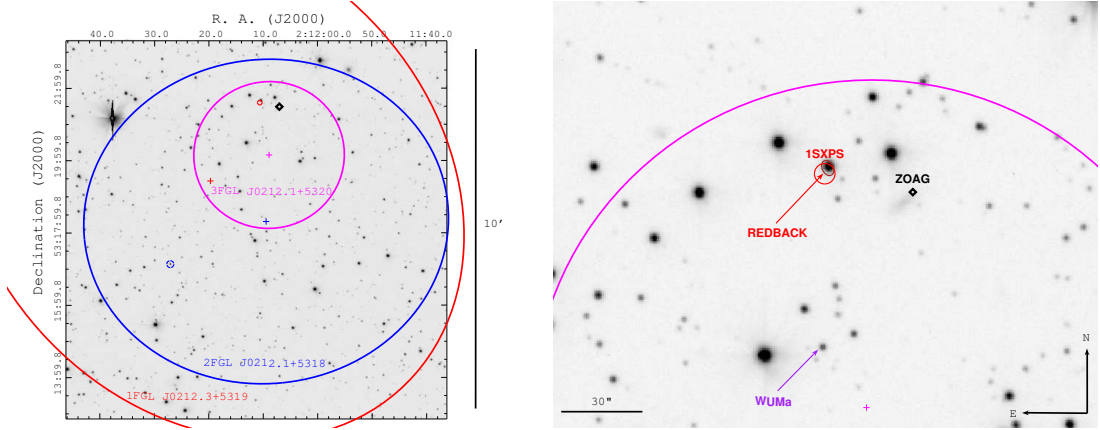}}}
%  \resizebox{1.0\columnwidth}{!}{\rotatebox{-90}{\includegraphics[]{f1b.eps}}}
%
  \caption{
{\it Left:} Full IAC-80-CAMELOT r' band image of the field of \fgl,
showing the (95\%) error ellipses from each of the three {\it Fermi}-LAT
point source catalogs, as indicated.
{\it Right:} Zoomed finding chart showing the location of our variable
optical source and redback candidate J0212 (P$_\mathrm{orb}
\simeq$20.9~hr, r'$\simeq$14.3mag; red arrow), the Swift X-ray
counterpart 1SXPS J021210.6+532136 (red circle), the nearby W~UMa
contact binary discovered in this work (P$_\mathrm{orb} \simeq$7.5~hr,
r'$\simeq$17mag, purple arrow; see Appendix~\ref{app:wuma}) and the
nearby galaxy ZOAG~G134.92-07.63 (black diamond). The brown ellipse
shows the Chandra 3-sigma location.
} %
    \label{fig:chart}
%\epsscale{1.0}
 \end{center}
\end{figure*}

Radio timing observations of {\it Fermi}-LAT sources have unveiled a
flurry of new pulsars \citep{Hessels11,Ray12}. However, black-widow
and redback MSPs are often occulted for a large fraction of the orbit
\citep{Archibald13}, making their direct detection as radio pulsars
challenging.
Blind searches for gamma-ray pulsations have met with some success
\citep{Pletsch12b}, yet they are computationally challenging,
especially when the signal is smeared out by Doppler shifts in short
(but unknown) orbital period binaries.

Here we take another approach to identify the {\it Fermi}-LAT source
\fgl, similar to that of \citet{Romani12} and \citet{Kong12}:
we search for and find a variable optical counterpart
(Section~\ref{sec:phot}) that matches a previously unidentified
X-ray source (Sec.~\ref{sec:xray}).
Our spectroscopic study (Section~\ref{sec:spec}) allows us to measure
the orbital period, the amplitude of the radial velocity curve, as
well as the companion's spectral type and projected rotational
velocity.
Together with the multi-wavelength properties of the source, which we
present in the rest of Section~\ref{sec:results}, this strongly
suggests that the binary hosts a recycled ``redback'' MSP.
We discuss the system's orbital parameters, potential and
peculiarities in Section~\ref{sec:discussion}.

\section{Data Analysis and Results}
\label{sec:results}

\subsection{Optical Photometry}
\label{sec:phot}

We observed the field of \fgl\ with the CAMELOT camera mounted on the
82~cm IAC-80 telescope, at the Teide Observatory.
As shown in Fig.~\ref{fig:chart}, the 10'$\times$10' field of view
covers all of the 2FGL and 3FGL location regions.
We observed \fgl\ in three epochs, 2014 August, 2015 February and 2015
December, obtaining 1--3 minute-long exposures with the SDSS g'r'i'
filters.
On 2015 February 14 we also calibrated the field against two
photometric standards, in photometric conditions.
The resulting photometric data set is summarized in Table~\ref{table:data}.

\subsubsection{Variability search and source identification}

We used the 2014-08-02/03 and 2015-12-11 observations to search for
variable counterparts, which had the longest uninterrupted sequences
of r' images lasting about 2 and 7 hours, respectively.
We identified 1296 objects in the field with signal-to-noise ratio
$\ge10$ and performed circular aperture photometry using the package
{\sc phot-iraf} and an aperture radius equal to the average full width at
half maximum (FWHM) of each image.
We carefully selected nine stars that remained stable
and used them as reference stars to perform differential photometry on
the remaining objects in the field.
We then measured the standard deviation ($\sigma$) and the average
value of the differential magnitude ($\Delta\,m$) from the light curve
of each object.
We estimate that our search for variability is sensitive down to a r'
magnitude of $\sim$20, while our faintest detected sources had
magnitudes of nearly 22.

We found two strongly variable objects, with $\sigma\gtrsim$70~mmag,
much more variable than the other objects at similar $\Delta\,m$
(which typically show $\sigma$=[1--3]~mmag).
We identify the most variable star (with a peak-to-peak light curve
amplitude $\Delta$g'$\simeq$0.4~mag) as a W~UMa-type contact binary in
the line of sight, as detailed in Appendix~\ref{app:wuma}.
This is shown with a purple arrow in Fig.~\ref{fig:chart} (right).
We also found a much less variable object ($<$14~mmag~hr$^{-1}$)
outside the error circle of \fgl\ (4.1' SE, about two times the 95\%
error radius), which we deem unrelated to the gamma-ray source.
This is shown with a dashed small blue circle in
Fig.~\ref{fig:chart} (left).

We find that the second most variable object, with
$\Delta$g'$\simeq$0.2~mag, is coincident with an X-ray source (red
circle and arrow and brown ellipse in Fig.~\ref{fig:chart}).
Based on its optical, X-ray and multi-wavelength properties, we argue
that this is the counterpart to \fgl\ and a ``redback'' binary MSP.
Using our astrometry-corrected 2015-12-11 r'-band images, we locate
the newly identified source at R.A.=02$^h$12$^m$10.46$^s$,
DEC=+53$^\circ$21$'$38.6$''$ (J2000), with an 0.4$''$ error radius
(FWHM/2).
Hereafter, we refer to this variable optical counterpart to \fgl\ as
simply J0212.

\begin{figure}
%\centering
  \begin{center}
  \resizebox{1.0\columnwidth}{!}{\rotatebox{0}{\includegraphics[]{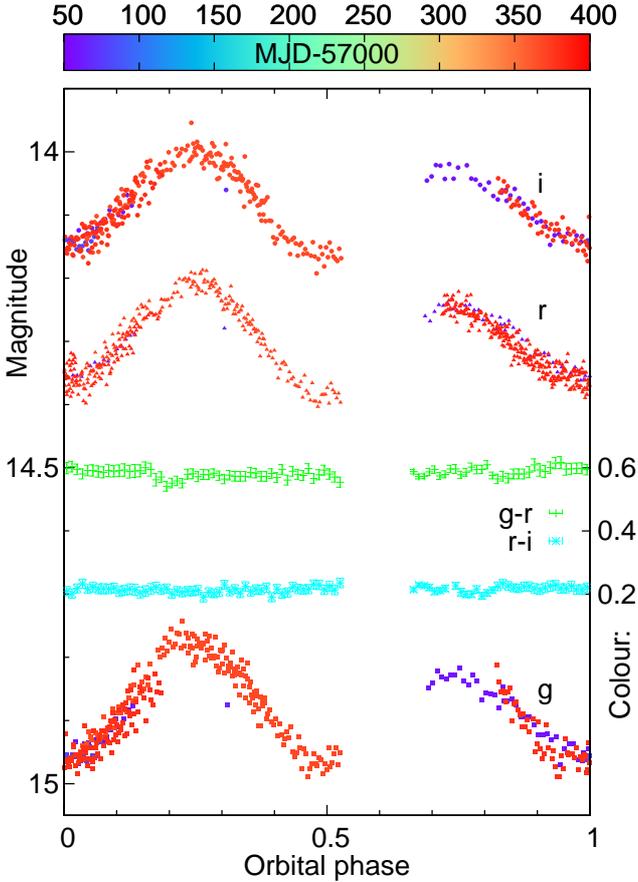}}}
  \caption{
Optical g'r'i' light curves of J0212 from our two photometric epochs
(February and December 2015; MJD date shown with colour scale, as
indicated), folded at the orbital period (0.86955~d).
The green and cyan symbols show the (g'-r') and (r'-i') colours,
respectively, with the scale indicated along the right axis.
The optical light curves are clearly asymmetric and show no
significant colour changes.
} %
    \label{fig:lc}
%\epsscale{1.0}
 \end{center}
\end{figure}

\subsubsection{Light curves of the redback MSP candidate}

We performed differential photometry of our variable object J0212 and
a nearby stable comparison star, using an aperture radius of 1.5-1.7
times the seeing .
We calibrated the g'r'i' magnitudes of the comparison star with
observations of the standard Hilt 233 \citep{Smith02} taken on the
same photometric night (2015 February 14) and using the colour and
extinction coefficients given by the IAC80-CAMELOT
team\footnote{http://www.iac.es/telescopes/pages/es/inicio/utilidades.php\linebreak\#camelot-calibracion}.

The light curves of J0212 in all three bands show qualitatively
similar variability: smooth broad asymmetric maxima and minima (see
Fig.~\ref{fig:lc}).
Besides this variability, indicative of orbital modulation in
semi-detached compact binaries
\citep[e.g.,][]{Avni75,Shahbaz96,Orosz97,Breton13}, J0212's light
curves are stable on time-scales of days to months.
We fitted the February-December 2015 light curves with a sine function
and obtained a photometric period P$_\mathrm{phot}$$\simeq$0.4348~d,
consistent with P$_\mathrm{orb}$/2 (see Section~\ref{sec:spec} below
for details).
The average magnitudes we find are 14.89, 14.30 and 14.08 in the g',
r' an i' bands, respectively.

\begin{table*}
%\footnotesize
\scriptsize
\caption{Summary of optical observations of 3FGL~J0212.1+5320 (J0212).}
\begin{minipage}{\textwidth}
\centering
\begin{tabular}{c c c c c c c c}
\hline\hline
%
%\footnotetext{}
Telescope & Instrument & Band\footnote{Photometric filters (effective wavelengths in \AA: g'=4639, r'=6122, i'=7439) or wavelength range covered by the spectra, in \AA.} & Date & Time & Exposures & Airmass & Seeing/Resolution \\
(diameter) & (configuration) & (filt., \AA) & (evening) & (UT) & (nr.$\times$duration) &  & (''/km~s$^{-1}$) \\
\hline
\multicolumn{8}{c}{\textbf{Photometry}}\\
\hline
IAC80-0.8m & CAMELOT-2x2 & r'     & 2014-08-02 & 04:09-05:37 & 15$\times$300s & 1.24-1.12 & 1.1-1.7''\\
IAC80-0.8m & CAMELOT-2x2 & r'     & 2014-08-03 & 04:30-05:20 & 10$\times$300s & 1.19-1.13 & 0.7-0.8''\\
IAC80-0.8m & CAMELOT-1x1 & g'r'i' & 2015-02-04 & 20:03-23:17 & 54$\times$180s & 1.17-1.90 & n.a.\\
IAC80-0.8m & CAMELOT-1x1 & g'r'i' & 2015-02-05 & 19:54-23:13 & 58$\times$180s & 1.16-1.91 & n.a.\\
IAC80-0.8m & CAMELOT-1x1 & g'r'i' & 2015-02-06 & 19:58-23:08 & 49$\times$180s & 1.17-1.90 & n.a.\\
IAC80-0.8m & CAMELOT-1x1 & g'r'i' & 2015-02-14 & 22:32-22:39 & 3$\times$180s & 1.86-1.92 & n.a.\\
IAC80-0.8m & CAMELOT-1x1 & g'i'   & 2015-02-26 & 20:13-21:47 & 28$\times$180s & 1.39-1.89 & n.a.\\
IAC80-0.8m & CAMELOT-1x1 & g'r'i' & 2015-12-10 & 19:43-03:04 & 229$\times$60s & 1.10-1.96 & n.a.\\
IAC80-0.8m & CAMELOT-1x1 & g'r'i' & 2015-12-11 & 19:41-02:57 & 289$\times$60s & 1.11-1.93 & 0.6-1.6''\\
IAC80-0.8m & CAMELOT-1x1 & g'r'i' & 2015-12-22 & 19:35-02:09 & 261$\times$60s & 1.10-1.90 & n.a.\\
IAC80-0.8m & CAMELOT-1x1 & g'r'i' & 2015-12-23 & 19:52-00:01 & 166$\times$60s & 1.10-1.31 & n.a.\\
IAC80-0.8m & CAMELOT-1x1 &     r' & 2015-12-28 & 19:30-01:51 & 257$\times$60s & 1.10-1.94 & n.a.\\
\hline\hline
\multicolumn{8}{c}{\textbf{Spectroscopy}}\\
\hline
WHT-4.2m & ACAM-VPH400 & 4100-9000 & 2015-12-16 & 19:07-20:49 & 6$\times$120s[2]\footnote{Very faint spectra were grouped by the numbers shown in brackets.} & 1.12-1.26 & 500\\
INT-2.5m & IDS-H1800V & 4400-5400 & 2015-12-23 & 23:59-01:01 & 6$\times$600s & 1.29-1.46 & 45 \\
INT-2.5m & IDS-R400V & 4000-7700 & 2016-01-19 & 20:32-00:22 & 54$\times$400s[2] & 1.13-1.64 & 155 \\
INT-2.5m & IDS-R400V & 4000-7100 & 2016-01-20 & 20:26-23:59 & 57$\times$200s[2] & 1.12-1.71 & 155 \\
INT-2.5m & IDS-R400V & 4000-7100 & 2016-01-21 & 19:45-00:01 & 28$\times$200s[2] & 1.10-1.75 & 155 \\
INT-2.5m & IDS-R400V\footnote{Using the EEV instead of the RED+2 CCD used in all other IDS spectra.} & 4000-7400 & 2016-01-26 & 20:46-23:06 & 18$\times$400s[3] & 1.18-1.52 & 155 \\
NOT-2.5m & ALFOSC-G7-1'' & 4000-6800 & 2016-02-03 & 20:39-00:04 & 24$\times$400s & 1.21-2.19 & 450 \\
INT-2.5m & IDS-R632V & 4200-6700 & 2016-02-25 & 20:56-23:10 & 12$\times$550s & 1.53-2.62 & 140 \\
NOT-2.5m & ALFOSC-G7-0.5'' & 4000-6800 & 2016-03-09 & 20:31-21:16 & 6$\times$500s & 1.63-1.94 & 240 \\
NOT-2.5m & ALFOSC-G7-0.5'' & 4000-6800 & 2016-03-14 & 21:31-22:16 & 6$\times$500s & 2.29-2.98 & 240 \\
NOT-2.5m & ALFOSC-G7-0.5'' & 4000-6800 & 2016-03-15 & 21:56-22:41 & 6$\times$500s & 2.69-3.68 & 240 \\
NOT-2.5m & ALFOSC-G7-0.5'' & 4000-6800 & 2016-03-16 & 21:36-22:22 & 7$\times$500s & 2.45-3.30 & 240 \\
\hline\hline
\end{tabular}
\end{minipage}
\label{table:data}
\end{table*}

\subsection{Optical Spectroscopy}
\label{sec:spec}

We obtained 230 medium-to-low resolution spectra of J0212 with the
William Herschel (WHT), Isaac Newton (INT) and Nordic Optical (NOT)
telescopes at the Roque de los Muchachos observatory, on La Palma.
The spectra were taken between December 2015 and March 2016, with
exposure times, central resolutions and instrumental setups summarized
in Table~\ref{table:data}.
We reduced these using standard \textsc{iraf} routines, extracted them
optimally with \textsc{starlink/pamela} \citep{Marsh89} and calibrated
the wavelength scale with interspersed arc lamp (Ne/Th/Cu/Ar) spectra,
fitting a polynomial to the wavelength-pixel relation (giving
residuals with rms more than 10 times smaller than the dispersion).
We then fine-tuned the wavelength scale taking the 5577.338~\AA\ and
6300.304~\AA\ sky emission lines as reference, thereby correcting for
sub-pixel offsets when present, and normalized the spectra dividing by
a spline fit to the continuum.

\begin{figure}
%\centering
  \begin{center}
  \resizebox{1.\columnwidth}{!}{\rotatebox{-90}{\includegraphics[]{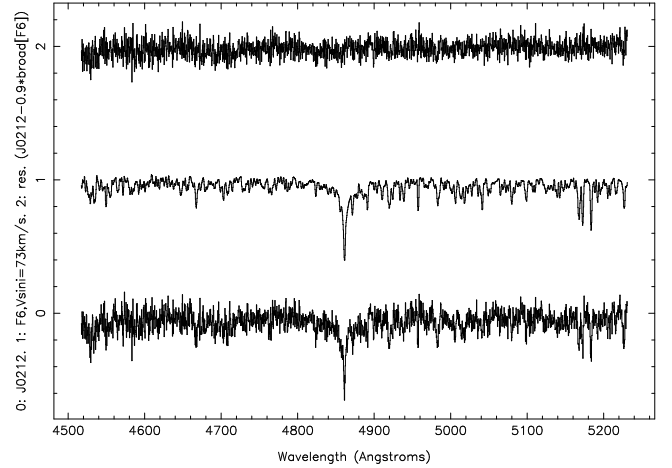}}}
  \caption{
Average normalized INT-IDS-H1800V spectrum of J0212 shifted to the
companion's reference frame (bottom), compared with an F6 template
spectrum broadened by Vsin(i)=73~km~s$^{-1}$ (middle).
Residuals from our best-match optimal subtraction are also shown
(top).
} %
    \label{fig:optsub}
%\epsscale{1.0}
 \end{center}
\end{figure}

\subsubsection{Spectral type and rotational velocity}

The spectra of J0212 feature hydrogen Balmer and metallic lines
typical of F-type main-sequence stars.
We applied the optimal subtraction method \citep{Casares93} in order to
measure the spectral type and temperature of J0212, comparing
quantitatively its photospheric absorption line spectra to a set of
UVESpop main sequence templates covering spectral types O--M
\citep{Bagnulo03}.
In all cases the spectra were shifted to the template rest frame to
remove the orbital velocity of J0212 (see below).
We find a spectral type of F6$\pm$2 from the 6 highest resolution
spectra taken on 2015-12-23 (which cover orbital phases 0.17--0.22)
using the hydrogen H$\beta$ (4861~\AA), MgI triplet (5167--5184~\AA)
and other metallic absorption lines in the 4500--5300~\AA\ range (see
Fig.~\ref{fig:optsub}).
Lower resolution IDS spectra taken around orbital phases 0.5 and 0.75
also indicate a F4--F8 spectral type, or an effective temperature
T$_\mathrm{eff}$=6640--6150~K.
Thus we do not find evidence for temperature changes along the orbit
larger than about 500~K.

Applying the same method with a set of template spectra broadened by
[0--200]~km~s$^{-1}$ in steps of 10~km~s$^{-1}$, we also measured the
projected rotational velocity of the companion star in J0212, Vsin(i).
Using F4, F6 and F8 templates we get consistent results for Vsin(i):
71.8$\pm$3.1~km~s$^{-1}$, 73.6$\pm$2.7~km~s$^{-1}$ and
73.9$\pm$2.6~km~s$^{-1}$, respectively (where the best value was
calculated from a parabolic fit to the lowest 3--5 points and the
errors correspond to $\Delta\chi^2$=1).
We adopt the weighted average of these three values as our final
measurement of Vsin(i)=73.2$\pm$1.6~km~s$^{-1}$.
The best-match template spectra were multiplied by a factor 0.7--0.9,
which suggests a non-stellar light veiling of 10--30\% in this band
(approximately equivalent to the photometric SDSS filter g').

\begin{figure}
%\centering
  \begin{center}
  \resizebox{0.8\columnwidth}{!}{\rotatebox{-90}{\includegraphics[]{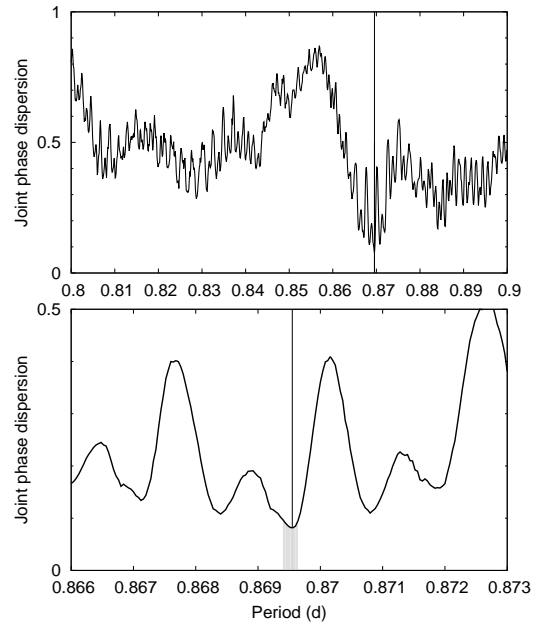}}}
  \caption{
{\it Top:} Phase dispersion diagram of J0212, using all available
observations (radial velocity and three-band light curves).
{\it Bottom:} Zoom into the strongest minimum at the orbital period
P$_\mathrm{orb}$=0.86955$\pm$0.00015~d, shown in gray and with the
vertical line (Sec.~\ref{sec:rvc}).
 } %
    \label{fig:pdm}
%\epsscale{1.0}
 \end{center}
\end{figure}

\begin{figure}
%\centering
  \begin{center}
  \resizebox{0.9\columnwidth}{!}{\rotatebox{-90}{\includegraphics[]{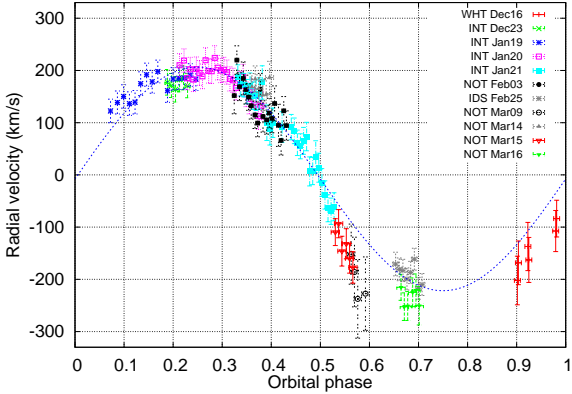}}}
  \caption{
Radial velocity curve of J0212, from cross-correlating the December
2015- March 2016 spectra of J0212 with an F6 template
(Sec.~\ref{sec:spec} for details).
The best-fit sinusoidal function with K$_2$=214.1$\pm$5.0~km~s$^{-1}$,
T0=57408.539$\pm$0.003 MJD(TDB),
and $\gamma$=-7.7$\pm$4.8~km~s$^{-1}$ is also shown.
} %
    \label{fig:rvc}
%\epsscale{1.0}
 \end{center}
\end{figure}

\subsubsection{Radial velocity curve}
\label{sec:rvc}

The rest of the spectra were cross correlated with a set of F
templates in order to measure radial velocities along the orbit,
keeping only cross-correlation values larger than 100 and adding a
systematic error of 20~km~s$^{-1}$ to the statistical errors to account for
residual uncertainties in absolute wavelength calibration.
We included the H$\alpha$ range (6513--6613~\AA) to calculate radial
velocities consistently (except in the 2015-12-23 spectra where only the
H$\beta$ range was available),
binned each series of spectra to match the corresponding dispersion
and broadened the templates in each case to match the spectral
resolution (Table~\ref{table:data}).

After applying the barycentric correction to both photometric and
spectroscopic data sets \citep{Eastman10}, we used the final set of 131
radial velocities and 629 r' band magnitudes in order to find and
measure the orbital period.
Due to nightly sampling and large data gaps, periodograms of the
radial velocity curve (RVC) and light curve (LC) are severly
affected by aliasing.
Using phase-dispersion minimization (PDM; \citealt{Stellingwerf78})
techniques, we find the strongest periodicity in the RVC at
0.8693$\pm$0.0004~d (errors quoted use a 20\% increase in PDM
statistics).
Although the RVC PDM diagram shows six other possible periods in the
0.8--0.9~d range, none of them produces a smooth folded RVC, and we
can rule them out.
% as artifacts.
%
%
To refine this period, we use the barycentred g', r' and i' band LCs,
which extend from February to December 2015.
The combined LC+RVC PDM diagram (see Fig.~\ref{fig:pdm}) shows the
strongest minimum at P$_\mathrm{orb}$=0.86955$\pm$0.00015~d,
consistent with and more precise than the period obtained from the RVC
only.

From a sine fit to the RVCs (Fig.~\ref{fig:rvc}), we measure a
semi-amplitude of K$_2$=214.1$\pm$5.0~km~s$^{-1}$,
a barycentric time of zero phase (companion at inferior conjunction)
T0=57408.539$\pm$0.003 MJD(TDB),
and a systemic velocity $\gamma$=-7.7$\pm$4.8~km~s$^{-1}$.
The errors reported here include the effect of using different
templates for the cross-correlation analysis, with spectral types
F4--F8.

\subsection{Gamma-ray}
\label{sec:gamma}

\begin{figure*}
%\centering
  \begin{center}
  \resizebox{1.6\columnwidth}{!}{\rotatebox{-90}{\includegraphics[]{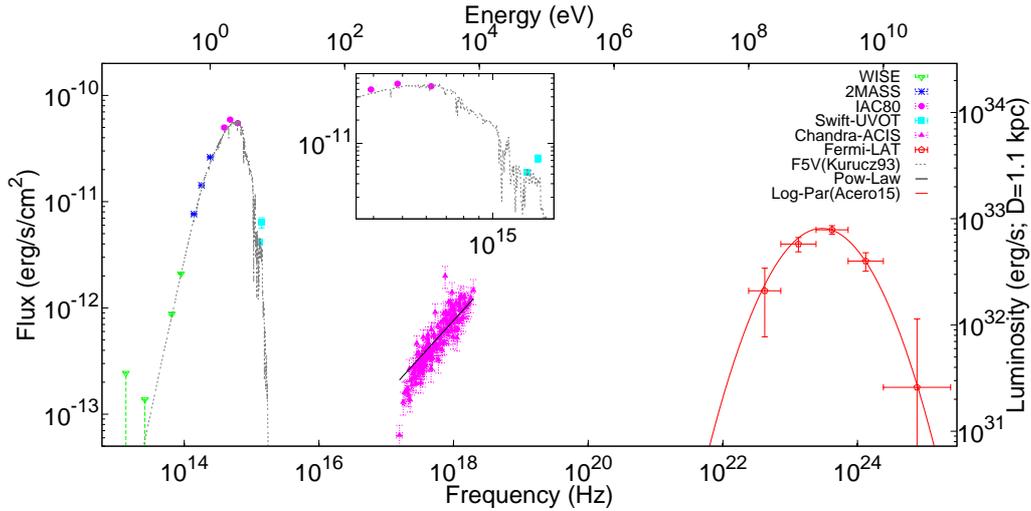}}}
  \caption{
Broadband spectral energy distribution of J0212, from IR to gamma-rays.
The optical fluxes correspond to phase 0 (Sec.~\ref{sec:discussion}),
and the X-ray data show the Chandra spectrum (Sec.~\ref{sec:xray}).
Fluxes from IR to UV are corrected for interstellar absorption.
For comparison, we show the spectrum of an F5V star with 1.2 R$_\odot$
radius at D=1.1~kpc (dashed gray line, not fitted; from
\citealt{Kurucz93}).
We also plot the best-fit power-law model (black line, corrected for
absorption; Sec.~\ref{sec:xray}) and the log-parabola fit to the LAT
spectrum (red line; from \citealt{Acero15}).
See Sections~\ref{sec:phot}--\ref{sec:IR} for details and references.
} %
    \label{fig:sed}
%\epsscale{1.0}
 \end{center}
\end{figure*}

As can be seen in Fig.~\ref{fig:chart}, the location of the
0.1-100~GeV gamma-ray source has improved with the increased exposure,
when going from the first (1FGL~J0212.3+5319) to the second
(2FGL~J0212.1+5318) and third (\fgl) {\it Fermi}-LAT catalogues
\citep{Abdo10,Nolan12,Acero15}.
The LAT location converges towards the position of J0212, our variable
optical counterpart (Sec.~\ref{sec:phot}).
Unless otherwise noted, we give in this section the most updated
values from the third {\it Fermi}-LAT source catalog.
The LAT error region is nearly circular with a 2$'$ radius, and the
spectrum is clearly curved \citep[log-normal in the description of][
  with significant curvature, 6.3, and a pivot energy of
  $\sim$1.1~GeV; see Fig.~\ref{fig:sed}]{Acero15}.

The LAT flux is relatively stable on timescales of weeks-months, with a
variability index of 51.5 (17.3 in the second catalog).
These characteristics (curved spectrum and low variability) are
similar to the known Fermi-LAT pulsar population
\citep[e.g.,][]{Abdo13}, which is why we selected \fgl\ for the
present work.
The integrated 0.1-100~GeV energy flux is
[1.71$\pm$0.16]$\times$10$^{-11}$~erg~cm$^{-2}$~s$^{-1}$
\citep{Acero15}.
The gamma-to-X-ray flux ratio ($\sim$10) is consistent with other
redback MSPs.
For the distance (D) derived in Section~\ref{sec:discussion}, the
corresponding gamma-ray (0.1-100~GeV) luminosity
L$_\gamma$=2.5$\times$10$^{33}$~(D/1.1~kpc)$^2$~erg~s$^{-1}$ is also
typical of MSPs \citep[most of which have
  L$_\gamma$$\sim$10$^{32}$--10$^{34}$~erg~s$^{-1}$;][]{Abdo13}.

\vspace{-0.2cm}

\subsection{X-ray}
\label{sec:xray}

{\it Chandra} (ACIS-S) observed the field on 2013 August 22 for a
total of 30~ks.
We find an X-ray point source at R.A.=02$^h$12$^m$10.50$^s$,
DEC=+53$^\circ$21$'$38.9$''$ (J2000), with an estimated 0.6$''$
uncertainty\footnote{While our work was being completed,
  \citealt{Parkinson16} tentatively identified this source,
  CXOU~J021210.5+532138, as a probable counterpart to \fgl.}.
As shown in Fig.~\ref{fig:chart} (brown ellipse), this is in excellent
agreement with the position of our variable optical counterpart,
J0212.
The background-corrected 0.5--10~keV spectrum is well fit
($\chi^2$/d.o.f.=134.3/142) with an absorbed power law model
(Fig.~\ref{fig:sed}).
We measure a photon index $\Gamma$=1.29$\pm$0.06, an absorbing
equivalent hydrogen column
N$_\mathrm{H}$=[1.4$\pm$0.3]$\times$10$^{21}$~cm$^{-2}$ and an
unabsorbed 0.5--10~keV flux of
[1.8$\pm$0.1]$\times$10$^{-12}$~erg~s$^{-1}$~cm$^{-2}$.
The background-corrected 0.3--10~keV count rate stayed roughly
constant at $\sim$0.1~c~s$^{-1}$ during the observation, which covered
about 40\% of an orbital cycle.
J0212 was also detected during {\it Swift}-XRT observations on 2010
October 9 and 12 for a total exposure of 4.5~ks, at a position and
X-ray flux consistent with the values reported above.

Taking as a reference the distance estimated in
Section~\ref{sec:discussion}, the 0.5--10~keV X-ray luminosity of
J0212 is
L$_\mathrm{X}$=2.6$\times$10$^{32}$~(D/1.1~kpc)$^2$~erg~s$^{-1}$.
This L$_\mathrm{X}$ and the very hard spectrum (photon index of 1.3) are fully
consistent with the rest of nearby redbacks in the pulsar state
\citep{Linares14c}.
Orbital X-ray variability from J0212 may be observed with longer
exposures, like that seen in other redbacks and interpreted as beaming
or partial occultation of the intrabinary shock
\citep[e.g.,][]{Bogdanov11b}.

\subsection{Ultraviolet}
\label{sec:UV}

{\it Swift}-UVOT observed the field in two occasions, on 2010 October
9 and 12, taking a 767s UV-M2 and a 1217s UV-W2 exposure, respectively.
We find a clear UV counterpart to J0212 in both filters;
using a 5$''$ aperture radius and the UVOT task {\sc uvotsource}, we
measure magnitudes of 17.99$\pm$0.05 and 18.16$\pm$0.09 in the W2 and
M2 filters, respectively.
The corresponding fluxes are shown in Fig.~\ref{fig:sed}, after
correcting for absorption (Sec.~\ref{sec:discussion}).

\subsection{Infrared}
\label{sec:IR}

Our variable optical counterpart matches a 2MASS source
\citep[2MASS~J02121047+5321387;][]{2mass} with the following
magnitudes: J=13.144$\pm$0.023, H=12.915$\pm$0.020 and
K=12.797$\pm$0.019.
% \citep{2mass,wise}
We also find a WISE source at the same location
\citep[WISE~J021210.46+532138.7;][]{wise}, with the following
magnitudes: w1=12.807$\pm$0.025, w2=12.759$\pm$0.026, w3$>$11.94 and
w4$>$9.16.
The corresponding fluxes are shown in Fig.~\ref{fig:sed}.

\section{Discussion}
\label{sec:discussion}

\subsection{Masses and orbital parameters}
\label{sec:m1}

We have discovered a variable optical counterpart to the gamma-ray
source \fgl, J0212, which coincides with a previously unclassified
X-ray source.
The multi-wavelength properties of J0212 are consistent with a binary
millisecond pulsar in a compact orbit (P$_\mathrm{orb}$=20.869[4]~h)
with a F6$\pm$2 main sequence companion star.
From the measured Vsin(i)=73.2$\pm$1.6~km~s$^{-1}$ and
K$_2$=214.1$\pm$5.0~km~s$^{-1}$ (Sec.~\ref{sec:spec}), assuming a
Roche-lobe filling, tidally locked and spherically symmetric companion
star, we find a mass ratio
q=M$_\mathrm{2}$/M$_\mathrm{1}$=0.26$^{+0.02}_{-0.03}$ \citep[where
  M$_\mathrm{2}$ and M$_\mathrm{1}$ are the masses of the
  secondary/companion and the primary/neutron star; see][]{Wade88}.
We note this is strictly a lower limit and thus q$\gtrsim$0.26, as the
companion may be smaller than its Roche lobe.

\begin{figure}
%\centering
  \begin{center}
  \resizebox{0.95\columnwidth}{!}{\rotatebox{-90}{\includegraphics[]{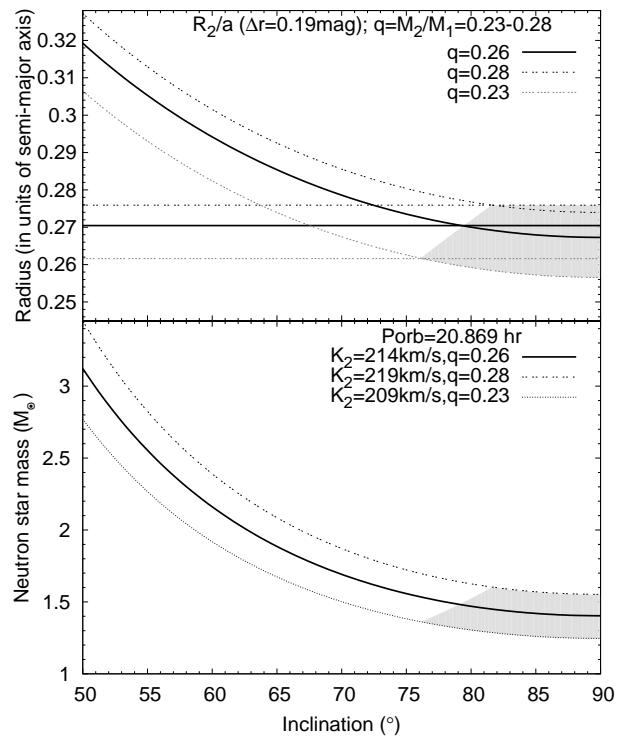}}}
  \caption{
{\it Bottom:} Mass of the primary or neutron star as a function of
inclination (i=90$^\circ$ corresponds to the orbital plane viewed edge
on). The curves shown use the indicated values of P$_\mathrm{orb}$,
K$_2$ and q (Sections~\ref{sec:spec} and \ref{sec:m1}).
{\it Top:} Radius of the secondary or companion star as a function of
inclination, in units of the semi-major axis a. Horizontal lines show
the Roche lobe radius R$_\mathrm{L2}$ for our estimated values of q,
as indicated. Curves show R$_2$/a for the observed light curve
amplitude $\Delta$r=0.19~mag \citep{Morris85}.
Gray-shaded regions show the preferred ranges of i, M$_1$ and R$_2$/a
from imposing that R$_2$$\leq$R$_\mathrm{L2}$ (Sec.~\ref{sec:m1}).
 } %
    \label{fig:m1}
%\epsscale{1.0}
 \end{center}
\end{figure}

These orbital parameters classify \fgl\ as a ``redback'' MSP
\citep[which have M$_\mathrm{2}$$\gtrsim$0.1--0.5~M$_\mathrm{\odot}$;
  e.g.,][]{Roberts11}.
The measured absorbing column density (Sec.~\ref{sec:xray})
corresponds to a pulsar dispersion measure DM$\sim$50~pc~cm$^{-3}$
according to the correlation presented in \citet{He13}.
This optical ephemeris will allow targeted searches for radio
(preferentially around phase 0.5 to avoid pulsar occultation) and
gamma-ray millisecond pulsations from this system, not reported to
date.
\fgl\ does not appear in the NRT radio pulsar search of {\it
  Fermi}-LAT sources presented by \citet{Guillemot12}, and we find no
radio counterpart in the NVSS 1.4~GHz survey \citep[][the field was
  not covered by FIRST]{Condon98}.

If its redback nature is confirmed, J0212 will have the longest
P$_\mathrm{orb}$ among the compact binary millisecond pulsars in the
Galactic field (both redbacks and black widows).
To our knowledge, only two redbacks with longer orbital period are
known, both residing in globular clusters: J1748-2446AD
(P$_\mathrm{orb}$=1.09~d) in Terzan 5 and J1740-5340
(P$_\mathrm{orb}$=1.35~d) in NGC~6397
\citep[][respectively]{Hessels06,DAmico01}.
From our K$_2$ and q measurements we derive the M$_1$-i relation shown
in Fig.~\ref{fig:m1} (bottom; using M$_1$=(1+q)$^2$ P K$_2$$^3$ /
(2$\pi$G sin$^3$(i))).
If the primary is indeed a neutron star, our results imply that
i$>$50$^\circ$ and M$_2$$<$0.8~M$_\odot$ for any plausible
M$_1$$<$3~M$_\odot$, and M$_1$$\gtrsim$1.3~M$_\odot$ for any i.

Furthermore, since we find that irradiation effects are negligible in
J0212 (Secs.~\ref{sec:spec} and \ref{sec:colours}), we can constrain
the inclination by ascribing the observed light curve amplitude
(peak-to-peak amplitude in r' $\Delta$r=0.19mag) to ellipsoidal
modulation of the tidally locked companion.
Using the analytical method presented by \citet[][equation (6) in
  particular]{Morris85} and taking limb- and gravity-darkening
coefficients for a Solar metallicity F5 star with log(g)=4.5
\citep{Claret11}, we can constrain the companion radius R$_2$/a
(Fig.~\ref{fig:m1}, top), where a is the semi-major axis of the
orbit.
Imposing that R$_2$ is smaller than the corresponding Roche lobe
radius \citep[R$_\mathrm{L2}$;][]{Eggleton83}, since there is no
evidence for mass transfer and accretion disk lines are not observed
(Sec.~\ref{sec:spec}), we find that i$\gtrsim$76$^\circ$ and therefore
M$_1$$\simeq$[1.3--1.6]~M$_\odot$ and
M$_2$$\simeq$[0.34--0.42]~M$_\odot$.
Thus according to the \citet{Morris85} relation, the inclination
should be high (i$\gtrsim$76$^\circ$) and the companion should be
close to filling its Roche lobe (R$_2$/R$_\mathrm{L2}$$>$98\%) in
order to produce the observed ellipsoidal modulation.

The M$_2$ constraints above imply that the companion is significantly
larger and hotter than an isolated star of its mass.
If we take M$_1$=1.5~M$_\odot$ and M$_2$=0.38~M$_\odot$
(i=80$^\circ$), the corresponding
R$_2$$\simeq$R$_\mathrm{L2}$=1.3~R$_\odot$ is roughly consistent with
the radius of an isolated F6V star.
Similar ``stripped'' or ``bloated'' companion stars, hotter and/or
larger than isolated stars of the same mass, are seen in redback
\citep{Crawford13} and black-widow MSPs \citep{Kerkwijk11}, as well as
neutron star transients in quiescence (\citealt{Bildsten01}; see also
\citealt{Orosz03} for further discussion).
Finally, we note that if the pulsar is detected, J0212 will be an
ideal system for an accurate neutron star mass measurement: it has a
bright, non-irradiated companion star in a likely high inclination orbit.
Pulsar timing and high-resolution spectroscopy can yield much more
precise measurements of q and K$_2$, respectively.
Detailed modelling of the optical light curve and spectral lines can
give tighter and more robust constraints on the inclination angle.

\subsection{Colours and broadband SED}
\label{sec:colours}

After correcting for interstellar reddening using E(B-V) =
N$_\mathrm{H}$ / (3.1$\times$1.8$\times$10$^{21}$~cm$^{-2}$) = 0.251
(with the N$_\mathrm{H}$ measured from the X-ray spectrum;
Sec.~\ref{sec:xray} and \citealt{Predehl95}), the corresponding
(g'-r') and (r'-i') colours are fully consistent with the F6 spectral
type we find from optical spectroscopy \citep[][see
  Sec.~\ref{sec:spec}]{Pecaut13}.
The infrared, optical and UV fluxes are also consistent with an F6 main
sequence star with radius $\simeq$1.3~R$_\odot$ at D$\simeq$1.1~kpc,
as shown in Fig.~\ref{fig:sed}.
We note that the W2 flux (the shortest wavelength UV measurement
available, at 2120~\AA) is about 40\% higher than the M2 flux (at
2310~\AA).
Comparing with stellar atmosphere models of a F5V star
(Fig.~\ref{fig:sed}; \citealt{Kurucz93}), we attribute this to a
relative drop in M2 flux due to FeII absorption bands in the
$\sim$2300--2400~\AA\ range.
The spectral energy distribution (Fig.~\ref{fig:sed}) also shows the
energy budget of the companion star ($\sim$10$^{34}$~erg~s$^{-1}$)
which dominates in the optical band, the gamma-rays from the putative
MSP ($\sim$10$^{33}$~erg~s$^{-1}$) and the shock between the MSP and
companion winds (intrabinary shock), which presumably powers the X-ray
emission ($\sim$10$^{32}$~erg~s$^{-1}$).

The colours remain approximately constant along the orbit
(Fig.~\ref{fig:lc}), implying little or no temperature change
between the different sides of the companion.
This complete lack of irradiation or ``heating'' of the companion by
the pulsar wind and radiation is exceptional among compact binary
MSPs.
From the allowed range of temperatures (6640--6150~K;
Sec.~\ref{sec:spec}), a semi-major axis of 4.7~R$_\odot$
(i=80$^\circ$) and assuming that the pulsar spin-down power $\dot{E}$
is emitted isotropically, we estimate an upper limit on
$\dot{E}$$<$[1--4]$\times$10$^{35}$~erg~s$^{-1}$ \citep[for an
  irradiation efficiency 10--30\%, following][]{Breton13}.
This limit is consistent with the $\dot{E}$ of most MSPs. Thus we
suggest that the lack of heating is simply due to the wide orbit of
J0212.
To our knowledge only PSR~J1740-5340, which is also in a long
P$_\mathrm{orb}$$\simeq$32~hr orbit, has shown a similar lack of
irradiation \citep{Orosz03}.

\subsection{Light curves and distance}

We measure phase-zero magnitudes of 14.96, 14.36 and 14.15 in the g',
r' an i' bands, respectively, which correspond to a dereddened V=13.83.
This makes J0212 the brightest compact binary MSP known to date
\citep[about two magnitudes brighter than PSR J1723-2837 in
  V,][]{Crawford13}.
For an F6 main sequence star with absolute magnitude
M$_\mathrm{V}$=3.7 \citep{Pecaut13}, implicitly assuming that the
companion radius is unperturbed, we estimate a distance to J0212 of
D=1.1$\pm$0.2~kpc (where the error corresponds to the allowed range of
spectral types, F4--F8).
As discussed in Section~\ref{sec:m1}, the radius we infer from our RVC
and LC analysis is consistent with this spectral type (but the mass is
not).

The optical orbital light curves are clearly asymmetric (Fig.~\ref{fig:lc}):
the light maximum at phase 0.25 (companion at ascending node) is 0.03,
0.04 and 0.06 magnitudes brighter than the maximum at phase 0.75
(descending node) in the i, r and g bands, respectively.
The minimum at phase 0 (companion at inferior conjunction) is about
0.01--0.03 magnitudes brighter than the minimum at phase 0.5.
While the asymmetry in the depth of the minima might be partly
explained by limb- and gravity-darkening effects, models for compact
binary MSP light curves in general, and their asymmetric maxima in
particular, are still under development
\citep{Breton13,Li14,Salvetti15}.
We leave detailed modelling for future work, and simply point out that
optical light curves similar to those of J0212 have been observed in
confirmed and candidate redback MSPs (PSR~J1628-32,
P$_\mathrm{orb}$$\simeq$5~hr, \citealt{Li14}; PSR~J2129-0429,
P$_\mathrm{orb}$$\simeq$15.2~hr, \citealt{Bellm16}; 1FGL~J0523.5-2529,
P$_\mathrm{orb}$$\simeq$16.5~hr?, \citealt{Strader14};
3FGL~J2039.6-5618, P$_\mathrm{orb}$$\simeq$5.4~hr?,
\citealt{Salvetti15}).

\subsection{X-rays and intrabinary shock}

J0212 features the highest X-ray luminosity
(2.6$\times$10$^{32}$~erg~s$^{-1}$; Sec.~\ref{sec:xray}) among redback
(in the pulsar state) and black-widow MSPs \citep{Linares14c}.
This places J0212 in the group of relatively X-ray luminous
redbacks. PSR~1740-5340, on the other hand, had $\sim$10 times lower
L$_\mathrm{X}$ \citep[][and references therein]{Bogdanov10}.
The wide orbit and lack of irradiation signatures strongly suggests
that the companion wind in these systems is not driven by MSP heating
effects.
Thus while irradiation of the companion appears to depend critically
on P$_\mathrm{orb}$, the luminosity of the intrabinary shock between
the pulsar and companion winds is not simply related to
P$_\mathrm{orb}$.
This may be due to a hotter companion with a larger mass loss rate in
the wind, which would compensate the larger orbital separation.

\subsection{Towards a systematic search}

To conclude, our results highlight the potential of small-aperture
optical telescopes like the IAC80 in identifying and characterizing
{\it Fermi}-LAT sources.
The exceptionally bright optical counterpart to \fgl\ that we have
discovered, with r'$\simeq$14.3~mag, sets the record for the brightest
MSP with a low-mass companion. A complete photometric survey of
unidentified Fermi-LAT sources should thus target a broad range
spanning more than 12 magnitudes, from the faintest g'$\sim$27~mag
black-widow counterparts \citep[e.g.,][]{Breton13} to the
g'$\sim$15~mag of J0212 or brighter.
It should be noted, moreover, that intra-night optical variability
searches are biased towards short-P$_\mathrm{orb}$ and strongly
irradiated systems, which give rise to drastic magnitude changes in
only a few hours \citep[e.g.,][]{Schroeder14}.
Our findings show that a systematic search for compact binary MSPs
should also target low amplitude ($\lesssim$0.05~mag~hr$^{-1}$) and
long period ($\gtrsim$12~hr) optical variability.

\textbf{Note:} Soon after our manuscript was submitted, Li et
al. (ArXiv:1609.02951) published a similar analysis of \fgl. Our
results mostly agree.

\textbf{Acknowledgments:}
This article is based on observations made with the IAC80, WHT, INT
and NOT telescopes operated on the islands of Tenerife and La Palma by
the IAC, ING and NOTSA.
We thank R. Alonso, R. Ashley, M. Balcells, P. Blay, P. Chinchilla,
E. G\'omez, A. Oscoz, C. Protasio and J. Telting for granting,
assisting or performing some of the observations presented here.
{\sc iraf} is distributed by the NOAO, operated by AURA under
cooperative agreement with the NSF.
We gratefully acknowledge the use of T. Marsh's {\sc molly}, {\sc
  pamela} and {\sc ultracam} analysis packages.
This research was supported by the Spanish MINECO under grants
AYA2013--42627 and AYA2012--38700.
PRG is supported by a Ram\'on y Cajal fellowship (RYC2010--05762).
JC acknowledges support by the Leverhulme Trust through the Visiting
Professorship Grant VP2-2015-046.
This publication makes use of data products provided by HEASARC, 2MASS
and WISE.

\vspace{-0.7cm}

%\bibliographystyle{mn2e}
%\bibliography{../biblio.bib}

\vspace{-0.6cm}

\appendix

\section{IAC80~J021210.8+532032.5: A new W UMa contact binary}
\label{app:wuma}

\begin{figure}
%\centering
  \begin{center}
  \resizebox{0.8\columnwidth}{!}{\rotatebox{0}{\includegraphics[]{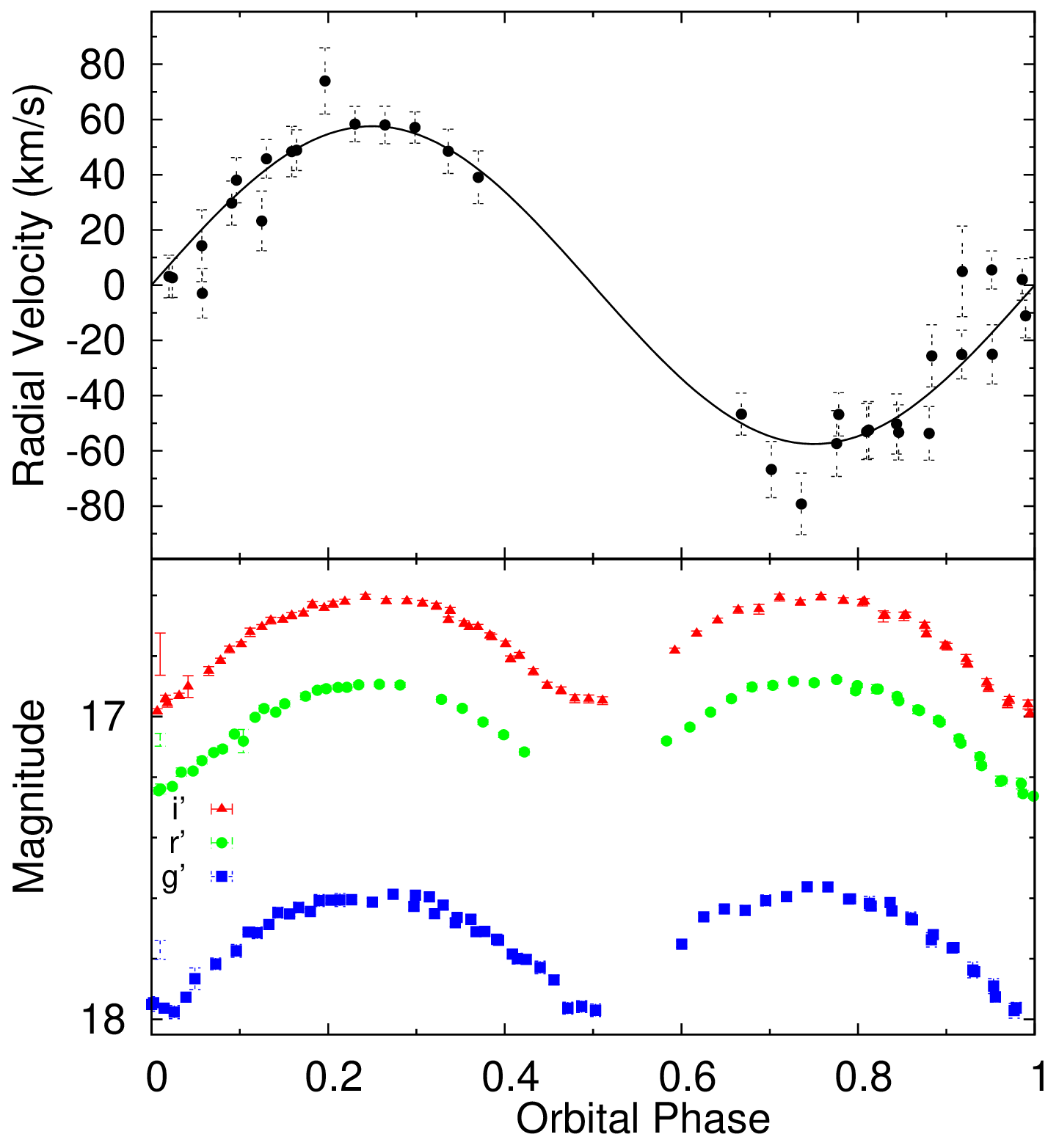}}}
%  \resizebox{1.0\columnwidth}{!}{\rotatebox{-90}{\includegraphics[]{f1b.eps}}}
%
  \caption{
{\it Top:} Radial velocity curve of J0212-WUMa measured with WHT-ISIS
in August 2015, folded at the orbital period
(P$_\mathrm{orb}$=7.469~h). The best-fit sine function is shown with a
solid line (systemic velocities removed).
{\it Bottom:} Phase-folded optical light curves of J0212-WUMa in three
bands, as indicated, measured with IAC80-CAMELOT in February 2015.
Error bars show the statistical error on the differential magnitude
and are typically smaller than the symbols, while average calibration
uncertainties are displayed along the left axis.
} %
    \label{fig:wuma-lcrv}
%\epsscale{1.0}
 \end{center}
\end{figure}

As a byproduct of our search, we report in this appendix the discovery
and fundamental properties of a new W Ursae Majoris (W UMa) system, an
eclipsing late-type contact binary system.
We name this object\footnote{Not listed in the following online W UMa
  catalogues: http://astro.utoronto.ca/\~rucinski/ogle.html ,
  http://www.oa-roma.inaf.it/maceroni/wumacat.html and
  http://www.konkoly.hu/staff/csizmadia/wuma.html}
IAC80~J021210.8+532032.5 and locate it at R.A.=02$^h$12$^m$10.77$^s$,
DEC=+53$^\circ$20$'$32.5$''$ (J2000), with an 0.4$''$ error radius
(FWHM/2).
The light curve folded at the orbital period
(Fig.~\ref{fig:wuma-lcrv}) shows broad maxima around phase 0.25 and
0.75, and narrower minima (flat in some cases, indicative of eclipses)
around phase 0.5 and 1.
We do not detect colour variations along the orbit, and constrain any
changes in g'-r' and r'-i' to be smaller than $\sim$0.05~mag.
These are all typical properties of W UMa-type binaries.

On 2015 February 27, we obtained seven 10-min low-resolution
($\sim$500~km~s$^{-1}$) WHT-ACAM-V400 spectra of
IAC80~J021210.8+532032.5, covering the 4500-9400~A range.
We reduced and extracted the spectra using standard procedures in {\sc
  iraf} and {\sc pamela}, including subtraction of the bias level,
flat-fielding and wavelength calibration with arc lamps fine-tuned
with sky lines.
The ACAM spectra of IAC80~J021210.8+532032.5 (taken during orbital
phase 0.15--0.42) show hydrogen Balmer (H$\alpha$, H$\beta$), sodium
(NaI), calcium (CaII), iron (FeI) and magnesium (MgI) absorption
lines, with little or no variability in intensity and width over the
1.2~h observation.
%
%As shown in Fig.~\ref{fig:acam}, 
%
These spectral features are typical of a G0 main sequence star, and we
infer a F5-G5 spectral type using the same techniques described in the
main text.

In order to measure the radial velocity curve, we obtained 33
medium-resolution WHT-ISIS spectra on 2015 August 26 and 27 with
the R1200 gratings.
The resulting spectra, extracted using the same procedures described
in Section~\ref{sec:spec}, have an average dispersion of
$\sim$0.5~\AA/pixel and a resolution of $\sim$50 km~s$^{-1}$
for the blue arm (4600--5400~\AA; because the sharper metallic lines
are present mostly in the blue arm, we focus our analysis on this
wavelength range).
Cross-correlation with a G0 template spectrum reveals the radial
velocity curve of IAC80~J021210.8+532032.5, with an orbital period
P$_\mathrm{orb}$=0.311$\pm$0.001~d (twice the photometric period), a
relatively low semi-amplitude (K$_2$=50.8$\pm$2.8~km~s$^{-1}$), a negative
systemic velocity ($\gamma$=-55.5$\pm$1.7~km~s$^{-1}$) and a barycentric
epoch of zero phase (inferior conjunction) T0=57262.125$\pm$0.002~MJD.
Fig.~\ref{fig:wuma-lcrv} shows both the light and radial velocity
curves folded at the orbital period.

\end{document}